\documentclass[conference]{IEEEtran}
\IEEEoverridecommandlockouts

\usepackage{pgfplots}
\usepackage{subcaption}
\usepackage{cite}
\usepackage{booktabs,makecell}
\usepackage{amsmath,amssymb,amsfonts}
\usepackage{algorithmic}
\usepackage{graphicx}
\usepackage{textcomp}
\usepackage{xcolor}
\usepackage{comment}
\usepackage{cleveref}
\usepackage{multirow}
\usepackage{makecell}

\def\BibTeX{{\rm B\kern-.05em{\sc i\kern-.025em b}\kern-.08em
    T\kern-.1667em\lower.7ex\hbox{E}\kern-.125emX}}
\begin{document}

\newcommand{\pr}{\mathbb{P}}
\newcommand{\nodes}{\mathcal{N}}              
\newcommand{\comp}{^{\mathrm{c}}}             
\newcommand{\ii}{\mathds{1}}   
\newtheorem{definition}{Definition}

\title{Improving the Robustness of the XRP Ledger Network via Edge Augmentation Strategies}

\makeatletter
\newcommand{\linebreakand}{%
  \end{@IEEEauthorhalign}
  \hfill\mbox{}\par
  \mbox{}\hfill\begin{@IEEEauthorhalign}
}
\makeatother

\author{
  \IEEEauthorblockN{Afonso Vilalonga}
  \IEEEauthorblockA{\textit{Universidade NOVA de Lisboa} \\
    \textit{NOVA LINCS}\\
    Portugal \\
    j.vilalonga@campus.fct.unl.pt}
  \and
  \IEEEauthorblockN{Orkun \.Irsoy}
  \IEEEauthorblockA{\textit{Electrical \& Computer Engineering} \\
    \textit{Carnegie Mellon University}\\
    Pittsburgh, USA \\
    oirsoy@andrew.cmu.edu}
  \and
  \IEEEauthorblockN{João S. Resende}
  \IEEEauthorblockA{\textit{Faculdade de Ciências} \\
    \textit{Universidade do Porto}\\
    Portugal \\
    jresende@fc.up.pt}
  \linebreakand 
  \IEEEauthorblockN{Henrique Domingos}
  \IEEEauthorblockA{\textit{Universidade NOVA de Lisboa} \\
    \textit{NOVA LINCS}\\
    Portugal \\
    hj@fct.unl.pt}
  \and
  \IEEEauthorblockN{Osman Ya\u{g}an}
  \IEEEauthorblockA{\textit{Electrical \& Computer Engineering} \\
    \textit{Carnegie Mellon University}\\
    Pittsburgh, USA \\
    oyagan@andrew.cmu.edu}
}

\maketitle

\begin{abstract}
The XRP Ledger allows its network participants to select a set of trusted peers within the network (i.e., the Unique Node List (UNL)) and communicate with them to reach consensus on which transactions should be included in the next ledger state. However, its consensus protocol requires significant overlap among participants’ UNLs, along with a high agreement threshold among the nodes within each UNL (e.g., 80\%). Consequently, an attacker could disrupt the consensus process in such a network by targeting the nodes that form the network’s connectivity backbone and reducing the number of trusted participants that can communicate with one another below the required threshold. In this paper, we evaluate a type of strategy to improve the robustness of the XRP Ledger’s existing topology, as measured by our formal definitions of quorum and network robustness, and compare it to a second strategy from prior work. The strategy we present is an addition/augmentation approach, in which new edges are added based on different constructions. The second strategy, originating from prior work, is a rewiring or edge-replacement approach, in which the overall number of edges is preserved but they are rearranged. For each strategy, we consider two different cases: one in which all nodes participate in the edge construction or rewiring process, and another in which only a subset of nodes participates. Our findings demonstrate substantial improvements in robustness when augmentation strategies are used over the default XRP Ledger topology and show that some augmentation strategies achieve robustness metrics equal to or exceeding those of prior work (i.e., rewiring-based approaches), even when the number of edges added is \emph{small}, e.g., three edges per node on average. Additionally, as part of a cost analysis, we show that the random K-out-based augmentation strategy maintains a higher topological similarity to the original network than rewiring, as measured by Jaccard similarity.

\end{abstract}

\begin{IEEEkeywords}
Network Robustness; Blockchain Security; {Random K-out Graphs}; XRP Ledger.
\end{IEEEkeywords}

\section{Introduction}
In blockchain networks, transactions must be proposed, validated, and committed to the chain through a consensus protocol, a process by which independent, potentially unreliable nodes agree on a single, coherent system state. One of the most widely used protocols in blockchains is the Nakamoto consensus~\cite{nakamoto2008bitcoin}, which relies on Proof of Work (PoW). In PoW, miners compete to solve a resource-intensive cryptographic puzzle, with the first to solve it committing a set of transactions to the ledger. A different popular alternative is Proof of Stake (PoS)~\cite{ethereum_pos2025}, in which validators are selected to propose and validate new blocks in proportion to the cryptocurrency they stake, replacing computational effort with economic commitment.

A key limitation of these consensus protocols is that certain nodes can exert greater influence over the ledger, not because the network inherently trusts them, but because of their superior computational or economic resources. Quorum-based consensus protocols take a different approach, achieving agreement when a quorum of participants approves the next set of transactions~\cite{malkhi1998byzantine}. Traditional Byzantine quorum-based protocols assume symmetric trust, where all participants are considered equally trustworthy~\cite{castro1999pbft}. In contrast, Federated Byzantine Agreement (FBA) protocols allow each participant to define its own trusted nodes (quorum slices) while also allowing open participation in the consensus process~\cite{Mazières2015SCP,xrpl_consensus_protocol}.

Consensus protocols inspired by or similar to FBA have been implemented in real-world blockchains. Examples include Stellar~\cite{Mazières2015SCP} and the XRP Ledger~\cite{XRPLedgerWebsite}. This work focuses on the XRP Ledger, which supports XRP, one of the largest cryptocurrencies by market capitalization~\cite{CoinMarketCap2025}. The XRP Ledger is a decentralized blockchain designed for high throughput, capable of processing thousands of transactions per second. To achieve consensus~\cite{xrpl_consensus_protocol}, nodes listen to their trusted validators, which are specialized nodes that participate in the consensus protocol, to determine the next ledger version. Each node maintains a static list of trusted validators, called the Unique Node List (UNL). Validators propose transaction sets and, through multiple rounds of consensus, agree on the correct set to commit. Nodes in the network declare that consensus is reached when at least 80\% of their trusted validators agree on the same set of transactions. If agreement falls below this threshold, validators adjust their proposals toward the majority view of the validators they trust and repeat the process until consensus is reached. To prevent network forks between nodes, the XRP Ledger recommends that all nodes' UNLs overlap by 90\%~\cite{xrpl_unl}.

The XRP Ledger consensus protocol’s requirement that 80\% of validators remain connected introduces a potential vulnerability, as it allows network disruption through partitioning or by disconnecting validators from one another, even when the validators themselves are not directly targeted; it is reasonable to assume that validators have hardened defense mechanisms due to their importance in the network. This threat is particularly concerning, as prior research~\cite{vytautas2023} has shown that a relatively small subset of nodes forms the backbone of the XRP Ledger’s connectivity. Targeted disruptions of these nodes could halt ledger progress, enabling attackers to censor blockchain usage or disrupt network operations.

Extensive research has been conducted on the XRP Ledger consensus protocol, ranging from analyses of required UNL overlap~\cite{Chase2018,klitos2020,ignacio2020} to formal security properties of the protocol~\cite{mauri2020formal}.
However, our focus is on network and quorum robustness, i.e., the ability of the XRP Ledger to remain connected and maintain consensus under adversarial conditions.
Tumas et al.~\cite{tumas2023federated} evaluated the robustness of the XRP Ledger against random node failures and targeted removals of high-degree nodes. Robustness was assessed using two metrics: network robustness and quorum robustness. Network robustness was defined as the percentage of nodes that must fail for the largest connected component to fall below half of the remaining active nodes. Quorum robustness was defined as the percentage of nodes that must be removed for fewer than 80\% of validators to remain connected. They showed that the XRP Ledger is robust to random failures, requiring the removal of 94\% and 84\% of nodes to compromise network and quorum robustness, respectively. However, under {\em targeted} attacks, only 20\% and 9\% of nodes needed to be removed to compromise these metrics. As a mitigation strategy, \cite{tumas2023federated} proposed modifying the network topology by {\em rewiring} existing edges in the network. Their approach~\cite[Algorithm 1]{tumas2023federated} is based on iterating through the network and removing or adding connections for each node to attain a specific ratio of {\em low}- to {\em high}-degree neighbors. This rewiring strategy enhances robustness against targeted attacks, but achieving the desired ratio may require numerous rewires per node, effectively approaching a redesign of the network topology and resulting in a topology differing significantly from the original.


In this work, we propose a different approach to improve the robustness of the XRP Ledger. Instead of rewiring existing edges, we propose adding new edges to the existing topology (i.e., an augmentation strategy), with new edges generated based on different graph construction models. Specifically, three different augmentation strategies are considered, based on three different graph construction models: random K-out (our primary construction strategy), random edge addition, and preferential attachment K-out. We discuss all three in greater detail in Section~\ref{methodology}. Our intuition for using augmentation strategies is based on prior work on graph construction models, specifically the random K-out model, which has shown that such graphs achieve almost-sure connectivity even at small scales (e.g., with $K=2$ and as few as 20 nodes, the probability of being connected exceeds 99.9\%~\cite{mansi_icc}). 

We implement both the rewire-based strategy and augmentation-based strategies on real XRP Ledger data and provide quantitative evidence of the comparison between the two, as well as between the different augmentation strategies we implement. Our results show that the K-out construction augmentation strategy is the best among those we develop, significantly improving both network and quorum robustness in the XRP Ledger and outperforming the rewiring approach~\cite{tumas2023federated} even under realistic operational constraints, such as limited node participation in the augmentation/rewire process and low $K$ values (e.g., 3 and 4). Additionally, we perform a cost analysis based on the similarity between the original and improved topologies, using Jaccard similarity. We show that the random K-out augmented network has a much higher similarity to the original topology than the rewired network. Beyond the specific application to the XRP Ledger, this paper introduces a new direction for the study of resilient network design: taking the union of a given network with an augmentation strategy such as the random K-out graph can be an effective mechanism for increasing its robustness without requiring a substantial redesign of the original topology. This insight opens up several promising avenues for future research, including analytical characterizations of robustness in such augmented topologies and empirical evaluations across other decentralized systems.

Summarizing, our main contributions are as follows: (1) We introduce a new augmentation-based strategy against targeted attacks on the XRP Ledger by augmenting its topology with edges generated using one of three graph construction-based augmentation strategies; (2) We compare the results of the three different augmentation strategies we implement; (3) We conduct a comparative analysis of the augmentation strategy that best performs against targeted attacks, in terms of quorum and network robustness (i.e., random K-out), against the rewiring strategy under realistic operational constraints; (4) We evaluate the topological differences between the original networks and the improved ones using Jaccard similarity; (5) We release an open-source implementation of our simulation framework~\cite{Vilalonga2025XRP}.

\section{Methodology}
\label{methodology}
In this section, we describe the methodology used to evaluate the network and quorum robustness of the XRP Ledger under the augmentation and rewiring strategies.

\subsection{Dataset}
We used the same dataset as in~\cite{tumas2023federated}, which was collected by scraping the XRP Ledger network hourly over a two-month period in 2022, comprising 1,290 snapshots. For each snapshot, we computed node and edge counts, degree statistics (average, minimum, maximum, and median), density, and average clustering coefficient, and then averaged these metrics across all snapshots. The snapshot used for our simulations was the one closest to these averages, based on Euclidean distance. Its statistics are as follows: 952 nodes, 15070 edges, an average degree of 31.7, a minimum degree of 1, a maximum degree of 342, and a largest connected component size of 952.

\subsection{Node Selection for the Attacks}
In line with~\cite{tumas2023federated}, our preliminary analysis has shown that the XRP Ledger is already highly robust against random failures. Therefore, we focus solely on targeted attacks, where nodes are selected based on either degree or betweenness centrality. Degree represents the number of connections a node has, while betweenness centrality measures how often a node lies on the shortest paths between all pairs of nodes. 
We also assume that attackers cannot directly target validators, which we consider a reasonable assumption, as validators may have stronger security defenses due to their critical role in the network. Moreover, directly targeting them would constitute a different type of attack from those analyzed in this work.

\subsection{Mitigation Strategies}
To improve the robustness of the XRP Ledger network, we propose augmenting its existing topology by adding new edges according to three different construction strategies. This allows us to evaluate which augmentation mechanism is most effective and to situate our primary proposal, the \textit{K-out augmentation} strategy, within a broader landscape of graph augmentation approaches. We also compare its effectiveness with the previously proposed \textit{rewiring} approach from~\cite{tumas2023federated}.

\textbf{Random edge addition:}
The existing topology is augmented by adding $K \cdot n$ new undirected edges, each placed uniformly at random from non-existing edges, where $n$ is the number of nodes and $K$ varies from 0 to 10 in increments of 1.
This mimics the edge-construction process of Erd\H{o}s--R\'enyi (ER) random graphs~\cite{erdos61conn,Bollobas,barabasi_book} and serves as a natural baseline, since ER graphs are known to be robust against targeted attacks when sufficiently dense~\cite{barabasi_book}.

\textbf{Preferential K-out:}
Each selected node establishes $K$ new undirected edges by connecting to $K$ peers chosen with probability proportional to their current degree, so higher-degree nodes are more likely to receive new connections.
This follows the preferential attachment mechanism of the Barab\'asi--Albert model~\cite{barabasi_book,barabasi}, and is a natural candidate to consider because the XRP Ledger's topology already exhibits scale-free characteristics consistent with preferential attachment mechanism~\cite{vytautas2023}.
Studying this strategy therefore provides intuition about the effect of continuing the network's natural growth trajectory on robustness.

\textbf{K-out augmentation:}
Each selected node establishes $K$ new undirected edges by connecting to $K$ peers chosen uniformly at random.
Unlike ER-based addition, where the total edge budget is spread across the network without any per-node guarantee, the K-out construction ensures that every participating node contributes exactly $K$ outgoing edges, producing a more uniform degree distribution in the added subgraph.
As shown in prior work~\cite{FennerFrieze1982,Yagan2013Pairwise,mansi_icc}, random K-out graphs achieve connectivity with high probability for any $K \ge 2$ using only $O(n)$ edges, compared with the $O(n \log n)$ edges required by ER graphs~\cite{erdos61conn,Bollobas}.
This efficiency advantage motivates its use as our primary augmentation proposal.

\textbf{Rewiring strategy:} In addition to the three augmentation strategies above, we compare against the rewiring approach from prior work. Following~\cite{tumas2023federated}, each selected node rewires its existing connections to adjust the ratio of low-degree (i.e., nodes with degree $< 100$) to high-degree (i.e., nodes with degree $\ge 100$) neighbors toward the ratio of 1:1, which was reported as optimal.
We apply 1, 5, 10, 15, 20, and 25 iterations of this rewiring procedure, with one rewire per node per iteration.
Our experiments indicate that the average critical attack size for both quorum and network robustness stabilizes after roughly 10 iterations. 

For all strategies, we also conducted experiments for cases where the improvements are applied to \textit{subsets} of nodes rather than the entire network to reflect realistic operational constraints, as some nodes in the XRP Ledger may be unable to accept new connections~\cite{tumas2023federated}.
Thus, we test subset sizes ranging from 20\% to 100\% of all nodes in increments of 20\%, where only nodes within the selected subset participate in either new edge generation or rewiring.


\subsection{Attack Simulations and Measuring Robustness}
We evaluate both robustness metrics through Monte Carlo simulations.
For each combination of $K$ value and subset size in our proposed mitigation strategy, we perform 300 independent repetitions to estimate the expected outcomes for each $(K, \text{subset})$ pair.
The same procedure is applied to the rewiring strategy, except that only the subset size varies, as no $K$ parameter is defined for that approach.

Because the dataset does not identify which nodes serve as validators, we construct validator sets following the same methodology as~\cite{tumas2023federated}. 
At the start of each simulation, 34 nodes are randomly selected to act as validators, matching the number in the recommended UNL at the time the dataset was collected. 
This random assignment also reflects a realistic scenario, as UNL lists may change over time. 
To assess the sensitivity of our results to this choice, we additionally consider two alternative policies: degree-proportional selection, which favors high-degree nodes, and inverse-degree selection, which favors low-degree nodes in selecting the validators. 

Full results are provided in Appendix~\ref{appendix_1}, where we show that the main conclusions remain consistent across all three assignment methods.

In each simulation, we compute each robustness metric by sequentially removing a fraction of nodes under targeted attacks, ranging from 0\% to 100\% in increments of 2\%, and record the results at each step. When the network reaches the failure threshold, defined according to the specific robustness metric being evaluated, the simulator restores the network to its initial state and proceeds to the next simulation. Between repetitions for a given parameterization, the subset of nodes selected for improvement, their target connections, and the validator set vary. Each of the 300 repetitions uses a unique random seed to ensure reproducibility across all $(K, \text{subset})$ configurations.

For both robustness metrics, we calculate the {\em critical} attack size, denoted \(p^*\). For network robustness, \(p^*_\text{network}\) represents the fraction of nodes that must be removed to fragment the network, as defined in Definition~1, while for quorum robustness, \(p^*_\text{quorum}\) represents the fraction required to halt the consensus protocol, as defined in Definition~2. Since $p^*$ is a random variable due to the randomness induced by the different topologies and chosen nodes, we report the {\em average} $p^*$ observed across all 300 repetitions for a specific robustness metric, $K$ and/or subset.

\smallskip
\noindent\textit{\textbf{Notation:}} Let \(G=(V,E)\) be the graph representing the XRP Ledger topology, where \(V\) is the set of nodes and \(E\) is the set of edges. Denote \(n := |V|\). Let \(S \subseteq V\) be the set of validators. Let \(\mathcal{C}(G) := \{C_1, \ldots, C_k\}\) be the set of connected components of \(G\). Let \(p \in [0,1]\) denote the attack size (i.e., the fraction of nodes removed). For a given attack of size \(p\), let \(\tilde{G}(p)\) denote the resulting graph after removals. 

\smallskip
\noindent\textit{\textbf{Definition 1: }}We define the {\em relative} size  of the largest connected component of \(\tilde{G}(p)\) as follows:
\[
\mathrm{LCC}(p)\;:=\;\frac{\max_{C\in\mathcal{C}\bigl(\tilde{G}(p)\bigr)} |C|}{n}
\]
The critical attack size at which the network is fragmented to below half of its original size is defined as
\[
p^*_{\textrm{network}} \;:=\; \inf \{\,p:\; LCC(p) < 0.5\,\}
\]

\smallskip
\noindent\textit{\textbf{Definition 2: }} Let \(Q \in [0,1]\) denote the quorum threshold (for the XRP consensus protocol, $Q=0.80$, corresponding to a minimum of 28 connected validators in a UNL of 34 validators). The 
critical  attack size at which the consensus protocol halts is defined as
\[
p^*_{\textrm{quorum}} \;:=\; \inf\left\{\,p:\; \frac{\max_{C\in\mathcal{C}\bigl(\tilde{G}(p)\bigr)} |C\cap S|}{|S|} < Q\,\right\}
\]
where $\max_{C\in\mathcal{C}\bigl(\tilde{G}(p)\bigr)} |C\cap S|$ gives the largest number of validators that are still in a connected component in the attacked network $\tilde{G}(p)$.




\section{Robustness Analysis}
We now present the results of our robustness analysis.

\subsection{Comparison of Augmentation Strategies}
\label{sec:aug_comparison}

Figure~\ref{fig:augmentation_comparison} compares all three augmentation strategies at full participation (100\% subset) against both attack types. The results show that random K-out augmentation outperforms both alternatives across all four robustness scenarios.

\begin{figure}[htbp]
    \centering
    \captionsetup[subfigure]{justification=centering} 

    \begin{minipage}[b]{0.47\columnwidth}
        \centering
        \includegraphics[width=\linewidth]{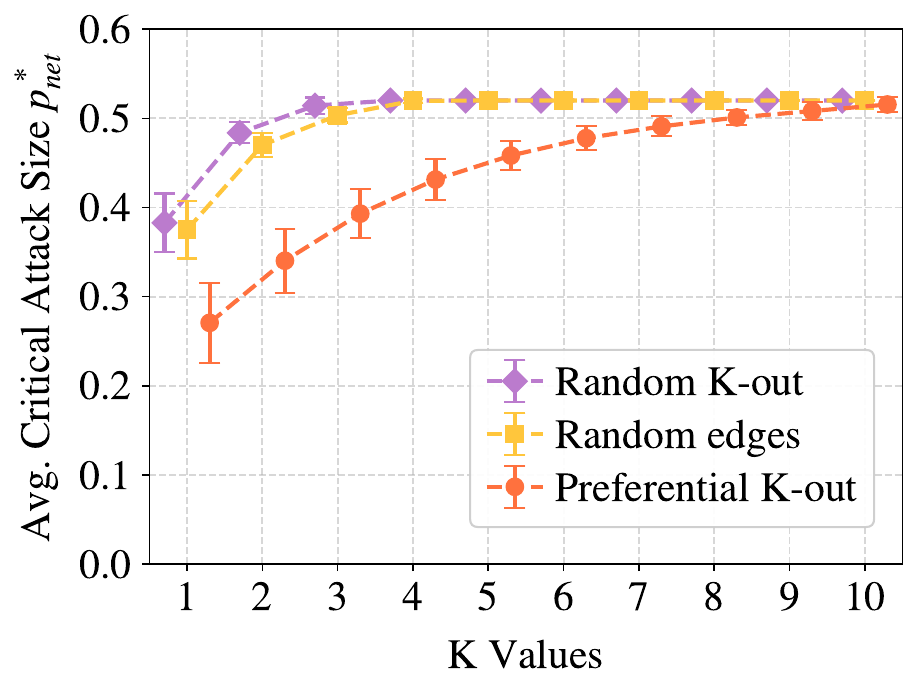}
        \subcaption{Network Robustness: High-degree attack.}
    \end{minipage}
    \hfill
    \begin{minipage}[b]{0.47\columnwidth}
        \centering
        \includegraphics[width=\linewidth]{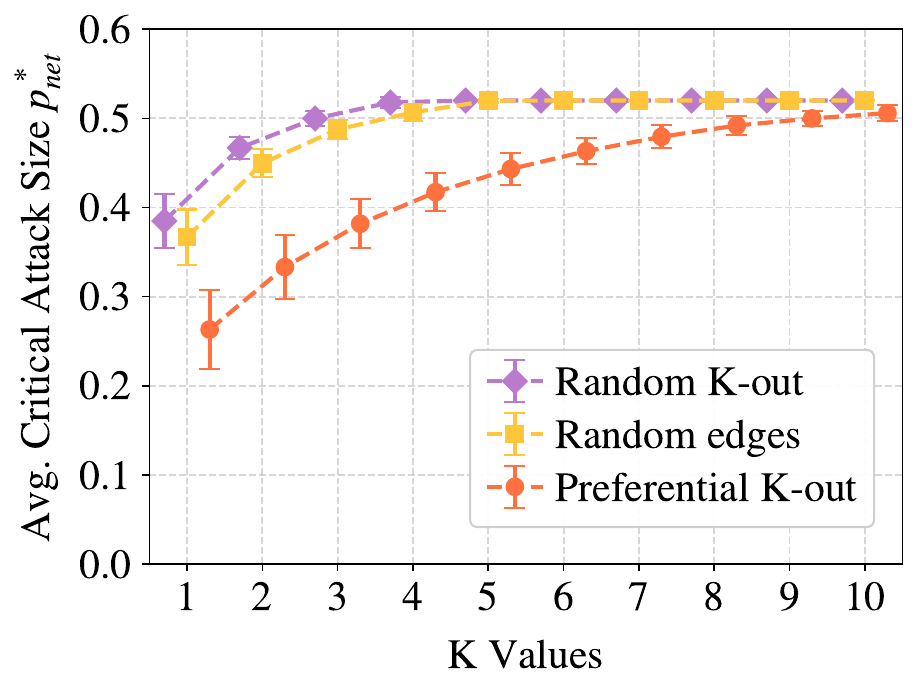}
        \subcaption{Network Robustness: High-betweenness attack.}
    \end{minipage}

    \vspace{0.35cm} 

    \begin{minipage}[b]{0.47\columnwidth}
        \centering
        \includegraphics[width=\linewidth]{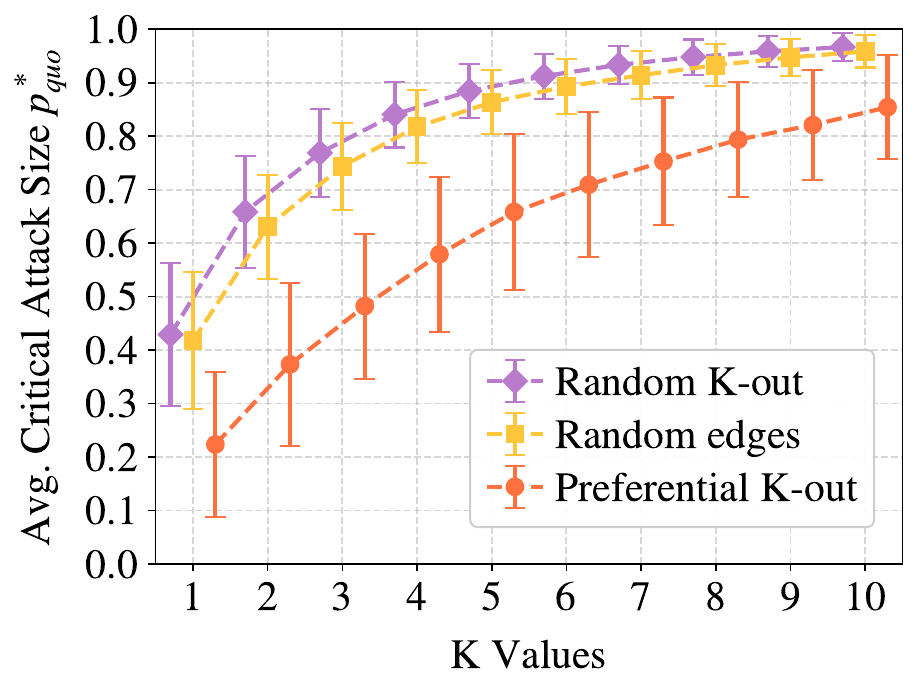}
        \subcaption{Quorum Robustness: High-degree attack.}
    \end{minipage}
    \hfill
    \begin{minipage}[b]{0.47\columnwidth}
        \centering
        \includegraphics[width=\linewidth]{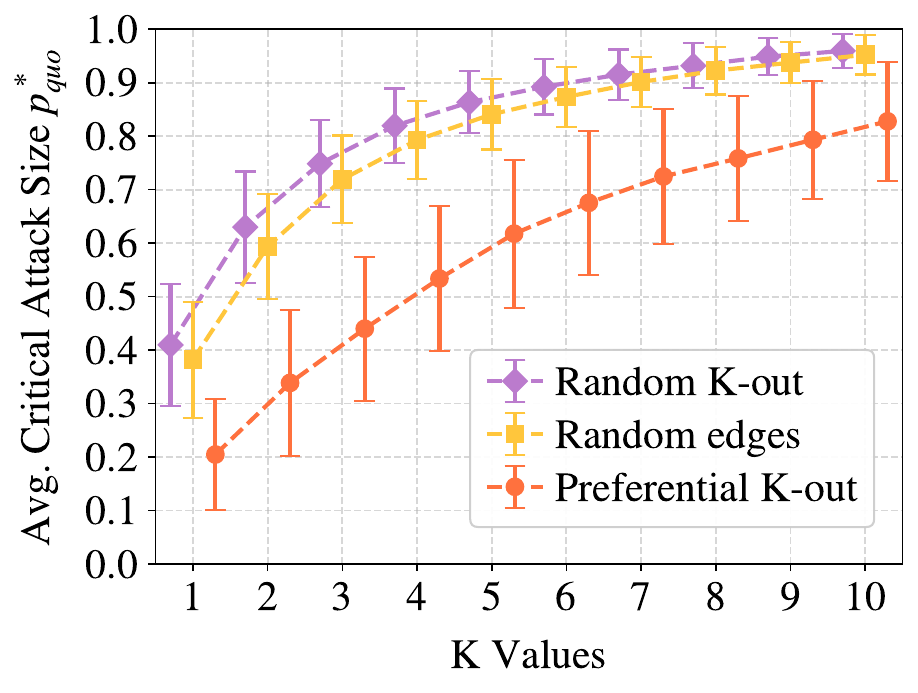}
        \subcaption{Quorum Robustness: High-betweenness attack.}
    \end{minipage}
    \caption{Targeted attack results for network and quorum robustness for the three augmentation strategies. \vspace{-2mm}}
    \label{fig:augmentation_comparison}
\end{figure}

Random edge addition provides solid improvements over the default topology, consistent with the known robustness of ER graphs to targeted attacks. However, it results in slightly worse robustness than K-out augmentation for the same $K$, particularly for quorum robustness. Preferential K-out performs noticeably worse than the other two strategies, especially at small $K$, and its gains in quorum robustness lag considerably even at $K=10$. This is expected: preferential attachment concentrates new edges on already high-degree nodes, which are precisely the nodes targeted in the targeted attack scenarios. Random K-out, by distributing new edges uniformly across all participating nodes, consistently achieves the best robustness among the three strategies.
Since K-out augmentation consistently outperforms the other two strategies, the remainder of the analysis focuses on comparing it against the rewiring approach across varying subset sizes. Results for the other augmentation strategies under different subsets are omitted for brevity.

\subsection{K-out Augmentation Strategy Applied to Subsets}

Our results for the K-out augmentation strategy indicate significant improvements in both network and quorum robustness, even for small values of $K$ (e.g., $K$ values up to 4) and feasible subset values (e.g., 40\% or 60\%). Specifically, to illustrate, for a subset size of 60\% with $K = 2$, the $p^*_\text{quorum}$ value increases from 11\% to 38\% under high-degree attacks (Figure~\ref{fig:augmentation}, (c) plot) and from 12\% to 33\% under high-betweenness attacks (Figure~\ref{fig:augmentation}, (d) plot). In other words, adding only two random connections per node to 60\% of the network nearly triples the quorum robustness, highlighting the effectiveness of the K-out augmentation strategy. Considering network robustness for the same setting ($K=2$, 60\%), $p^*_\text{network}$ increases from 19\% to 35\% and from 18\% to 34\% for high-degree and high-betweenness attacks, respectively (Figure~\ref{fig:augmentation}, (a) and (b) plots, respectively).

\begin{figure}[hbtp]
    \centering
    \captionsetup[subfigure]{justification=centering} 

    \begin{minipage}[b]{0.47\columnwidth}
        \centering
        \includegraphics[width=\linewidth]{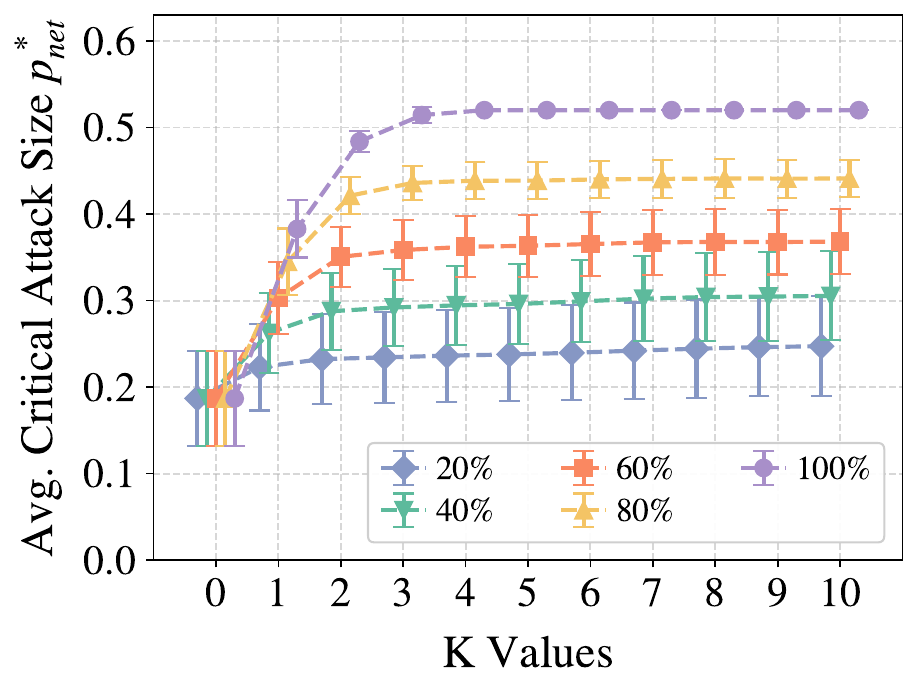}
        \subcaption{Network Robustness: High-degree attack.}
    \end{minipage}
    \hfill
    \begin{minipage}[b]{0.47\columnwidth}
        \centering
        \includegraphics[width=\linewidth]{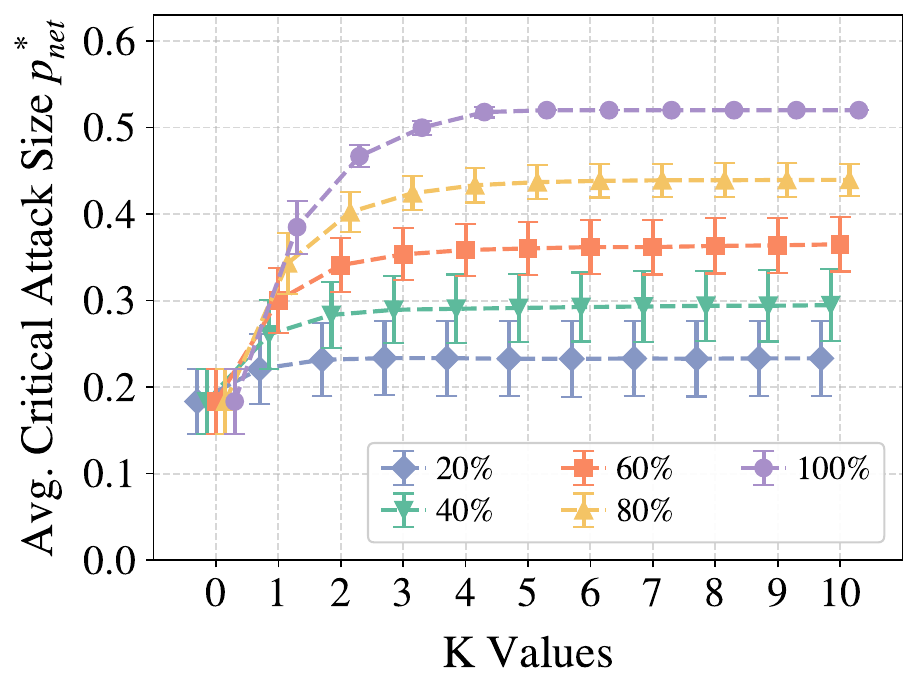}
        \subcaption{Network Robustness: High-betweenness attack.}
    \end{minipage}

    \vspace{0.35cm} 

    \begin{minipage}[b]{0.47\columnwidth}
        \centering
        \includegraphics[width=\linewidth]{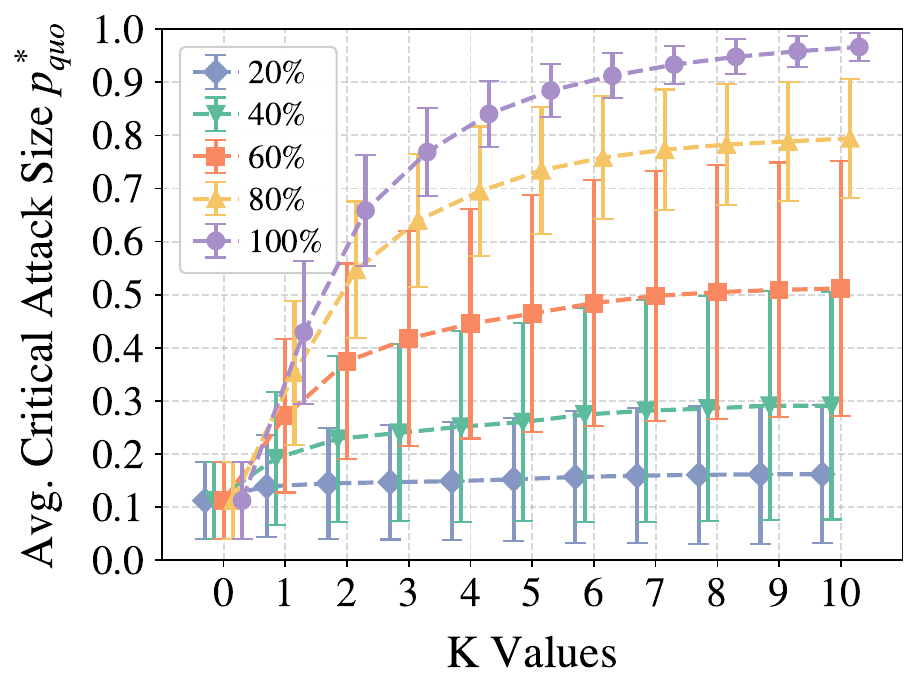}
        \subcaption{Quorum Robustness: High-degree attack.}
    \end{minipage}
    \hfill
    \begin{minipage}[b]{0.47\columnwidth}
        \centering
        \includegraphics[width=\linewidth]{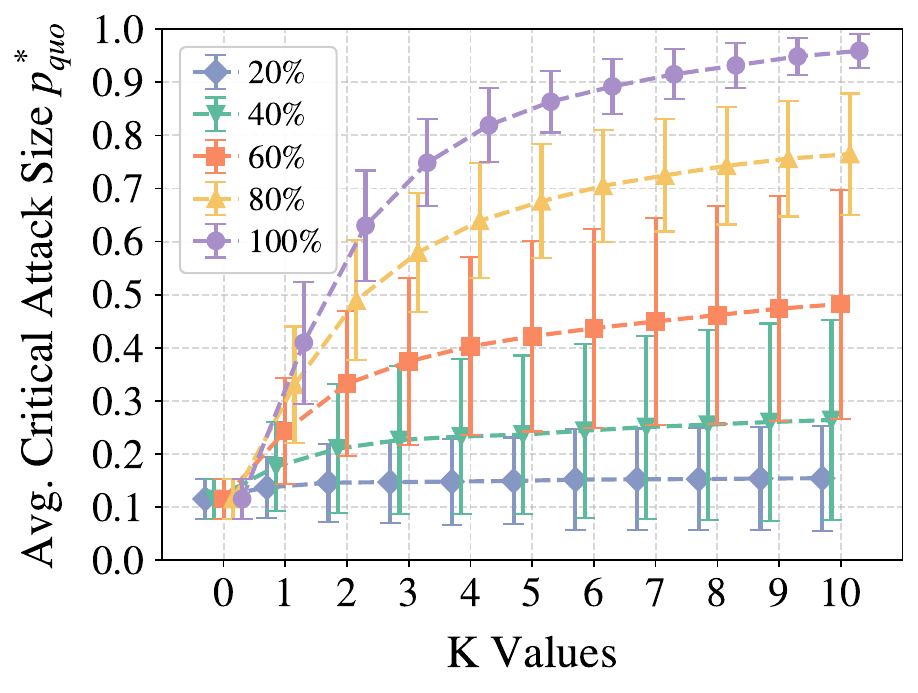}
        \subcaption{Quorum Robustness: High-betweenness attack.}
    \end{minipage}
    \caption{Targeted attack results for network and quorum robustness for the augmentation strategy. \vspace{-2mm}}
    \label{fig:augmentation}
\end{figure}

The results also indicate that the behavior of the plotted curves is directly influenced by both the subset size and the $K$ values. In smaller subsets, the differences between $K$ values are less pronounced, with plateaus reached at lower $K$ values than in larger subsets. By contrast, the 80\% and 100\% subsets show substantial differences between $K$ values, with plateaus starting at higher $K$ values. This behavior likely reflects the reduced impact of improving only a small percentage of nodes compared to larger percentages.

\subsection{Comparison Between the K-out Augmentation and Rewiring Strategies}
To compare the augmentation and rewiring strategies, we analyze the performance of the rewiring strategy for varying subset sizes. Figure~\ref{fig:rewiring} presents the resulting robustness curves for the rewiring strategy, where the horizontal axis represents the number of rewiring iterations and the vertical axis corresponds to the critical attack size. The rows represent network and quorum robustness, and the columns correspond to high-degree and high-betweenness attacks, respectively. Table~\ref{tab:results2} provides a complete comparison of the two methods for each targeted attack, robustness metric, and subset size. In the table entries, $I$ denotes the iteration at which the maximum $p^*$ is achieved by the rewiring strategy, while $K$ represents the minimum $K$ value at which the K-out augmentation strategy matches or exceeds the performance of the rewiring strategy.

\begin{figure}[!htbp]
    \centering
    \captionsetup[subfigure]{justification=centering} 

    \begin{minipage}[b]{0.47\columnwidth}
        \centering
        \includegraphics[width=\linewidth]{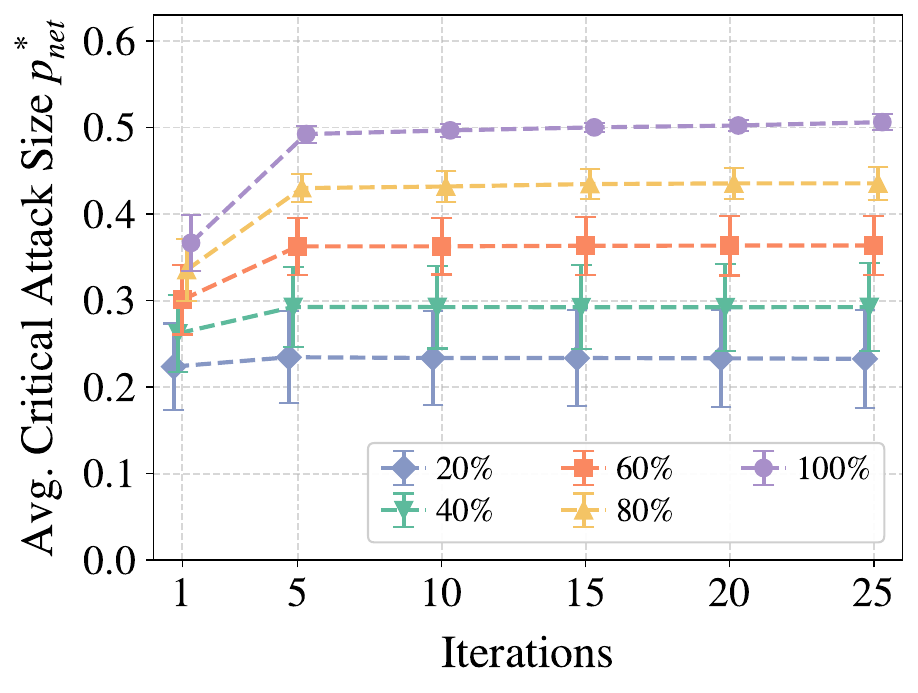}
        \subcaption{Network Robustness: High-degree attack.}
    \end{minipage}
    \hfill
    \begin{minipage}[b]{0.47\columnwidth}
        \centering
        \includegraphics[width=\linewidth]{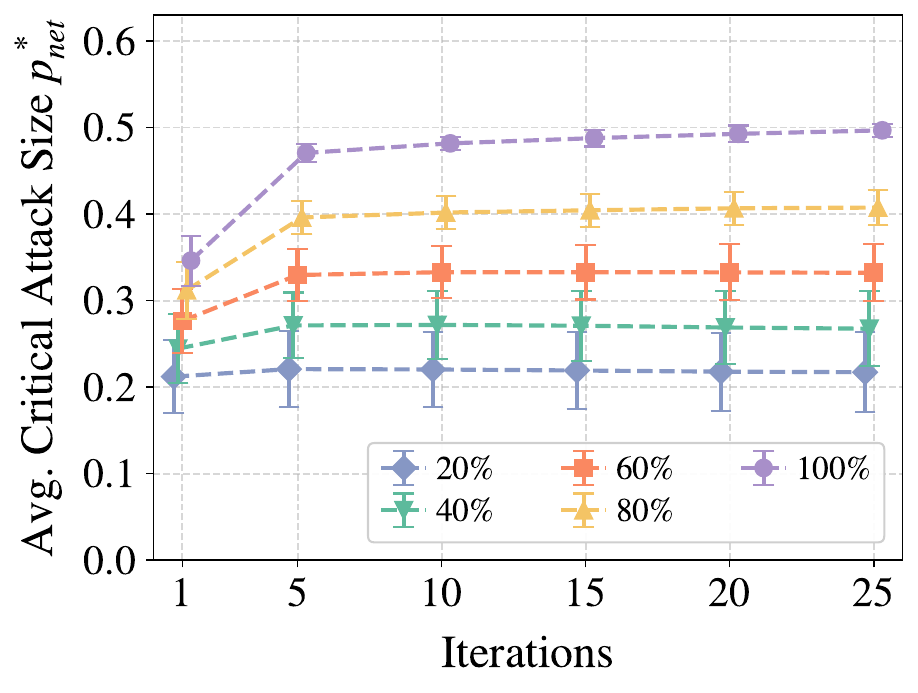}
        \subcaption{Network Robustness: High-betweenness attack.}
    \end{minipage}

    \vspace{0.35cm} 

    \begin{minipage}[b]{0.47\columnwidth}
        \centering
        \includegraphics[width=\linewidth]{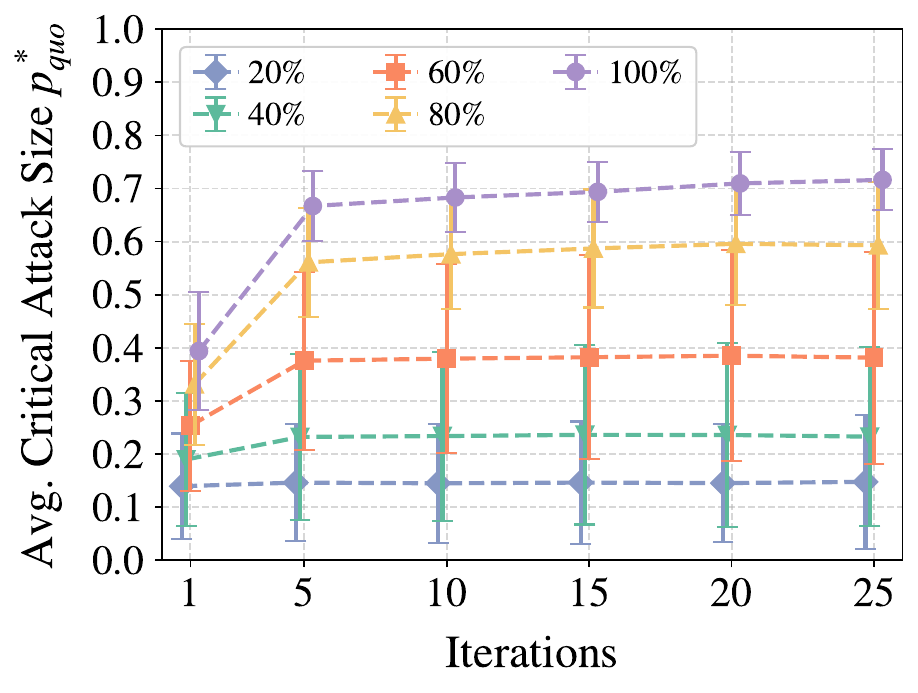}
        \subcaption{Quorum Robustness: High-degree attack.}
    \end{minipage}
    \hfill
    \begin{minipage}[b]{0.47\columnwidth}
        \centering
        \includegraphics[width=\linewidth]{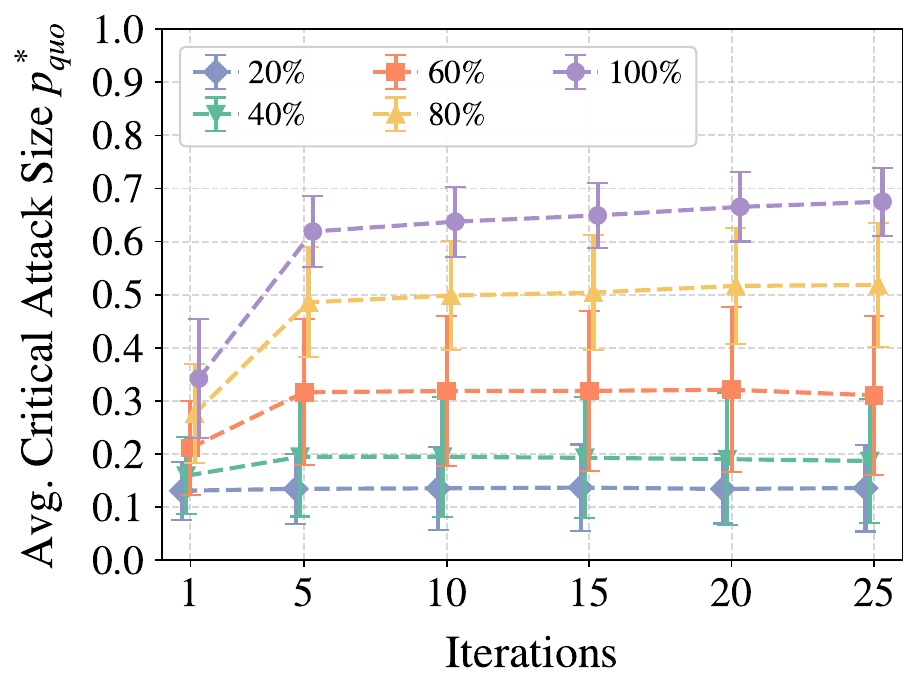}
        \subcaption{Quorum Robustness: High-betweenness attack.}
    \end{minipage}

    \caption{Targeted attack results for network and quorum robustness under the rewiring strategy.}
    \label{fig:rewiring}
    \vspace{-3mm}
\end{figure}

\begin{table}[!htbp]
\centering
\caption{The smallest $K$ values for which our mitigation achieves $p^*$ values equal to or greater than those obtained with the rewiring strategy for the same subset. (D) = degree; (B) = betweenness.}
\scalebox{0.9}{ 
\begin{tabular}{l@{\hskip 2pt} cc@{\hskip 10pt} cc}
\hline
\rule{-3pt}{2.5ex}
\multirow{2}{*}{\textbf{Subset}} 
  & \multicolumn{2}{c}{\textbf{Network ($K$, Iteration)}} 
  & \multicolumn{2}{c}{\textbf{Quorum ($K$, Iteration)}} \\
\cline{2-5}
 & \rule{0pt}{2.5ex}\textbf{D} & \textbf{B} & \textbf{D} & \textbf{B} \\
\hline
\rule{0pt}{2.5ex}20\% & $K=3$ $I=5$ & $K=1$ $I=5$ & $K=4$ $I=25$ & $K=1$ $I=15$ \\
40\% & $K=4$ $I=5$ & $K=2$ $I=10$ & $K=3$ $I=15$ & $K=2$ $I=10$ \\
60\% & $K=6$ $I=25$ & $K=2$ $I=15$ & $K=3$ $I=20$ & $K=2$ $I=20$ \\
80\% & $K=3$ $I=25$ & $K=3$ $I=25$ & $K=3$ $I=20$ & $K=3$ $I=25$ \\
100\% & $K=3$ $I=25$ & $K=3$ $I=25$ & $K=3$ $I=25$ & $K=3$ $I=25$ \\
\hline
\end{tabular}
}
\label{tab:results2}
\end{table}
The results in Table~\ref{tab:results2} indicate that, overall, the augmentation strategy can match or exceed the performance of the rewiring approach with relatively small $K$ values.
Considering both network and quorum robustness against high-betweenness attacks, $K = 3$ is sufficient to achieve or surpass the performance of rewiring for subset sizes of 80\% and 100\%, and even lower values are enough for smaller subsets.
A similar pattern is observed for high-degree attacks, except for a few cases, such as $K = 6$ for network robustness at a subset size of 80\%.
Nonetheless, even in this instance, for $K \ge 4$, the resulting $p^*$ value is nearly identical, differing only after the second decimal place between the augmentation and rewiring strategies.

\begin{figure}[hbtp]
    \centering
    \captionsetup[subfigure]{justification=centering} 

    \begin{minipage}[b]{0.47\columnwidth}
        \centering
        \includegraphics[width=\linewidth]{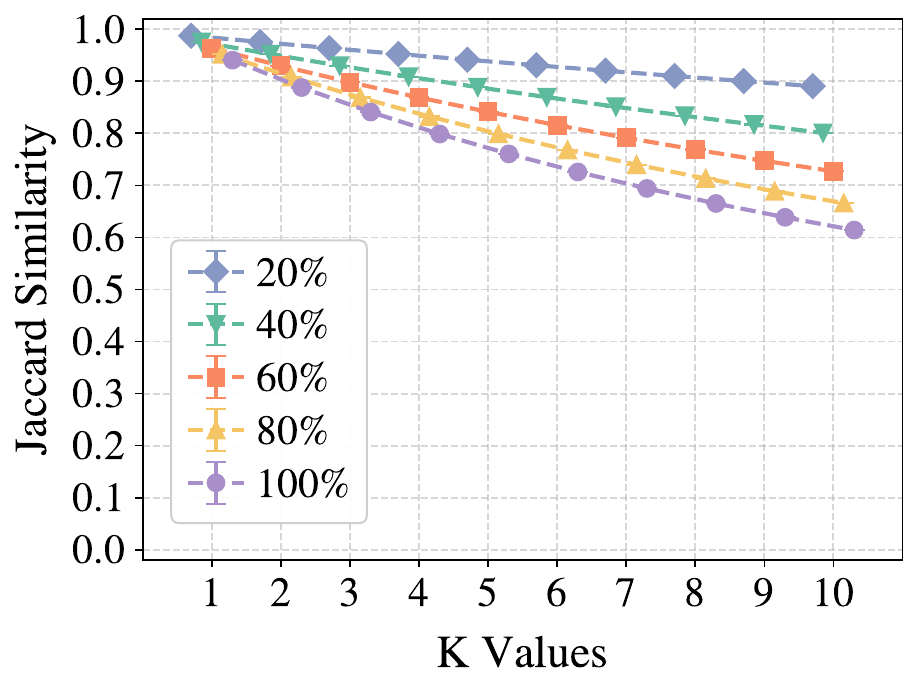}
        \subcaption{K-out Augmentation}
    \end{minipage}
    \hfill
    \begin{minipage}[b]{0.47\columnwidth}
        \centering
        \includegraphics[width=\linewidth]{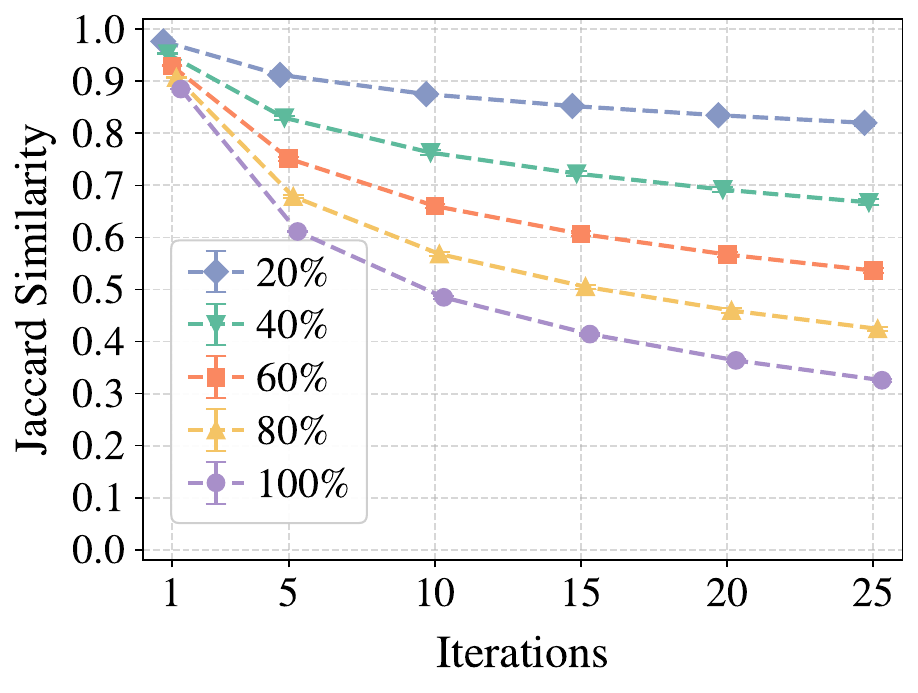}
        \subcaption{Rewiring}
    \end{minipage}
    \caption{Jaccard similarity between the original and improved edge sets for the K-out augmentation (left) and rewiring (right) strategies, across subset sizes. \vspace{-2mm}}
    \label{fig:Jaccard_comp}
\end{figure}

Beyond robustness performance, the two strategies also differ in how much they reshape the original topology. 
Figure~\ref{fig:Jaccard_comp} shows the Jaccard similarity between the edge sets of the original and improved networks for both strategies across subset sizes. 
From Table~\ref{tab:results2}, for 80\%-100\% subsets, $K=3$ augmentation matches the robustness gains of rewiring at around 20-25 iterations, yet the Jaccard similarity stays around 0.85 for augmentation while dropping well below 0.5 for rewiring. 
In other words, augmentation achieves the same robustness improvement while keeping nearly three times more of the original topology intact.

This points to a tradeoff worth considering. While rewiring preserves the total edge count, each operation involves both removing an existing connection and establishing a new one, fundamentally altering the original topology.
The two approaches also differ in deployment complexity: rewiring requires each node to inspect its neighbors' degrees over multiple iterations, demanding coordination and information sharing.
K-out augmentation needs none of this; each node independently picks $K$ random peers with no knowledge of the rest of the network (apart from the existence of the nodes it selects), making it easy to deploy in a decentralized setting.
Moreover, as the network grows, new nodes simply generate $K$ random connections upon joining, allowing robustness gains to be naturally maintained without additional coordination.

\section{Conclusion}
In this paper, we analyze the improvements resulting from applying different graph constructions to the XRP Ledger, focusing on their impact on network and quorum robustness, both of which we formally define. We compare the results of the implemented augmentation strategies among themselves and with prior research, both in cases where all nodes participate in the edge augmentation or rewiring process and in a potentially more realistic scenario in which only a subset of nodes participates in making outgoing and accepting incoming edges. Our findings show that augmentation strategies generally improve both robustness metrics compared with the default network, with the K-out augmentation strategy outperforming the rewiring strategy. While this paper focuses on the XRP Ledger, we believe these results suggest promising avenues for applying augmentation strategies to other decentralized networks and for studying their effects on robustness.

\section*{Acknowledgment}
We thank Dr. Vytautas Tumas
from Ripple for his thoughtful insights and discussions that helped shape
the initial direction of this paper. 
This work was supported in part by the Secure Blockchain Initiative at CyLab Security and Privacy Institute and by the Air Force Office of Scientific Research (AFOSR) Grant \# FA9550-22-1-0233. In addition, A. Vilalonga was supported by the FCT Ph.D. scholarship grant PRT/BD/154787/2023, awarded through the CMU Portugal Affiliated Ph.D. Program, and by UID/04516/NOVA Laboratory for Computer Science and Informatics with the financial support of FCT.IP.

\bibliographystyle{IEEEtran}
\bibliography{ref}

\appendices

\begin{figure*}
    \centering
    \begin{minipage}[t]{0.49\textwidth}
        \centering
        \textbf{Augmentation Strategy}\\[0.3em]
        \begin{subfigure}[b]{0.48\linewidth}
            \centering
            \includegraphics[width=\linewidth]{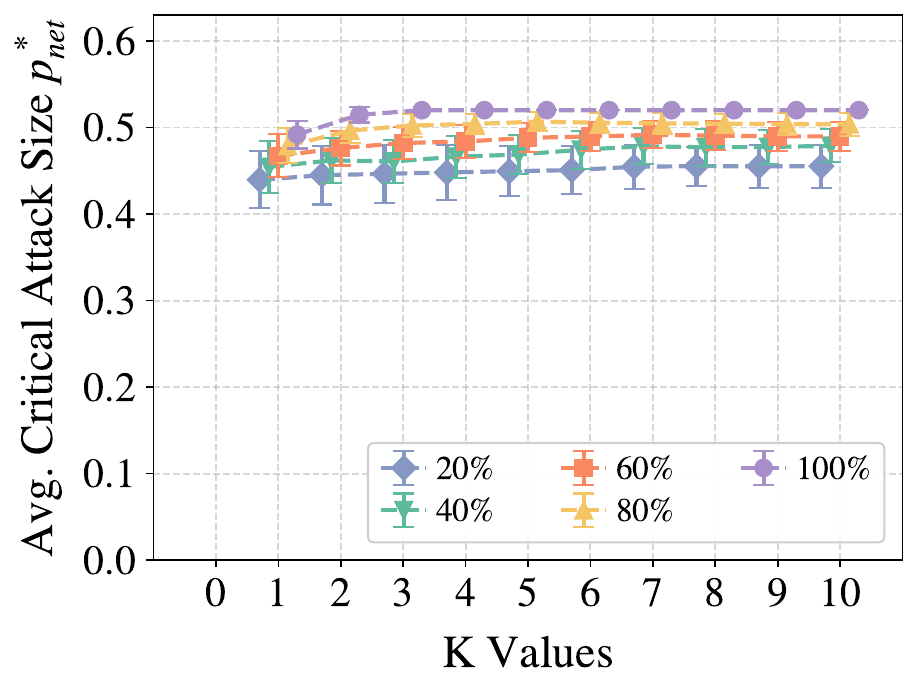}
            \caption{Network: degree attack}
        \end{subfigure}
        \hfill
        \begin{subfigure}[b]{0.48\linewidth}
            \centering
            \includegraphics[width=\linewidth]{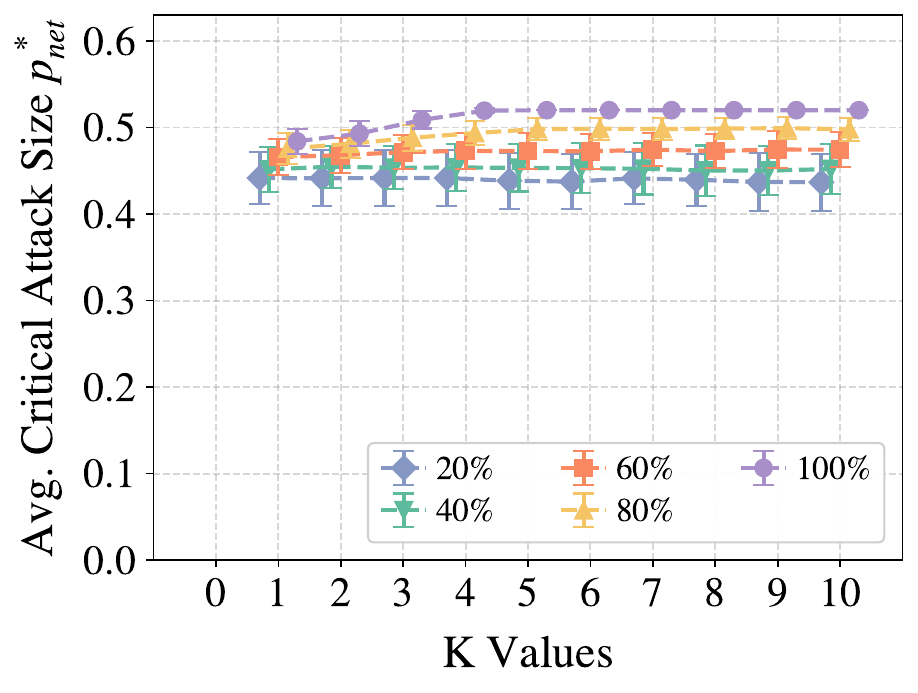}
            \caption{Network: betweenness attack}
        \end{subfigure}

        \vspace{0.25cm}

        \begin{subfigure}[b]{0.48\linewidth}
            \centering
            \includegraphics[width=\linewidth]{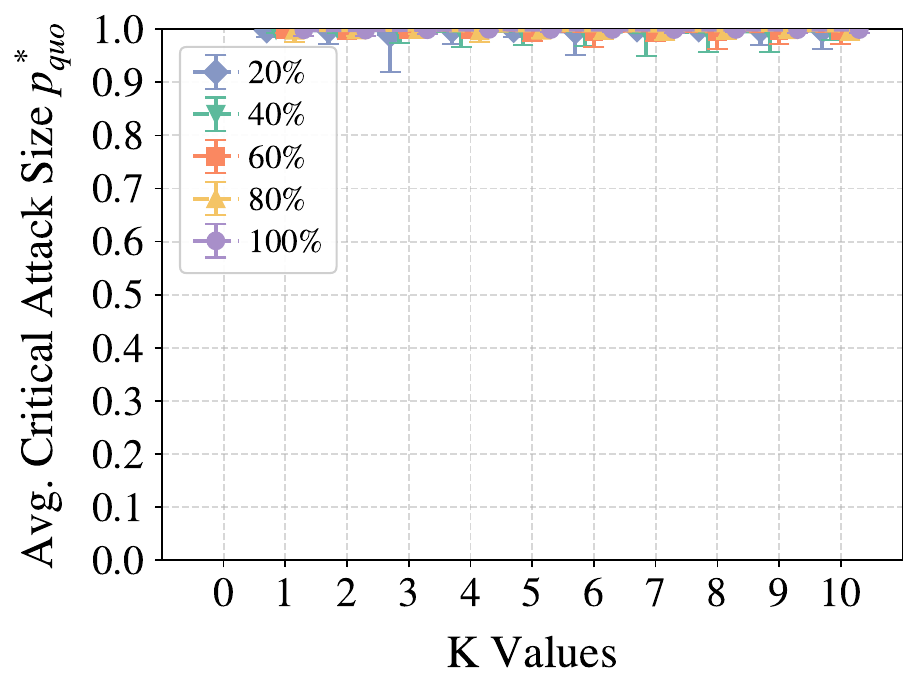}
            \caption{Quorum: degree attack}
        \end{subfigure}
        \hfill
        \begin{subfigure}[b]{0.48\linewidth}
            \centering
            \includegraphics[width=\linewidth]{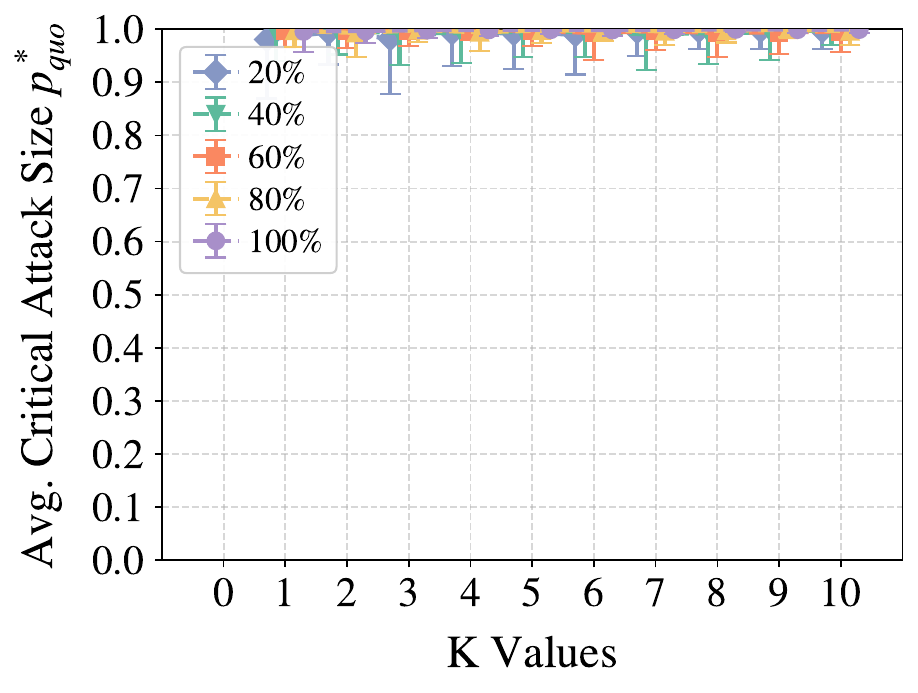}
            \caption{Quorum: betweenness attack}
        \end{subfigure}
    \end{minipage}
    \hfill
    \begin{minipage}[t]{0.49\textwidth}
        \centering
        \textbf{Rewiring Strategy}\\[0.3em]
        \begin{subfigure}[b]{0.48\linewidth}
            \centering
            \includegraphics[width=\linewidth]{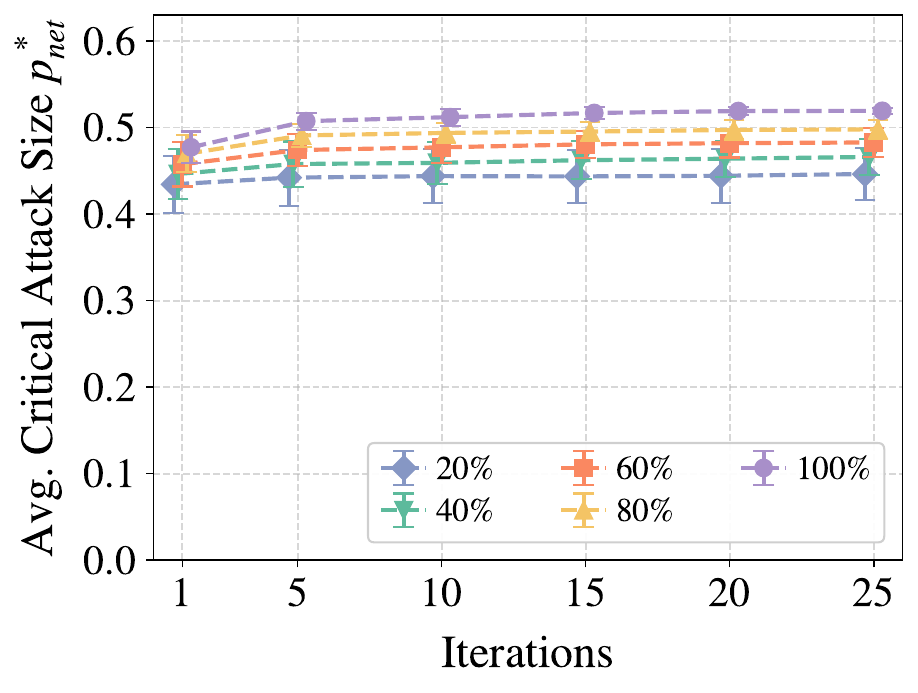}
            \caption{Network: degree attack}
        \end{subfigure}
        \hfill
        \begin{subfigure}[b]{0.48\linewidth}
            \centering
            \includegraphics[width=\linewidth]{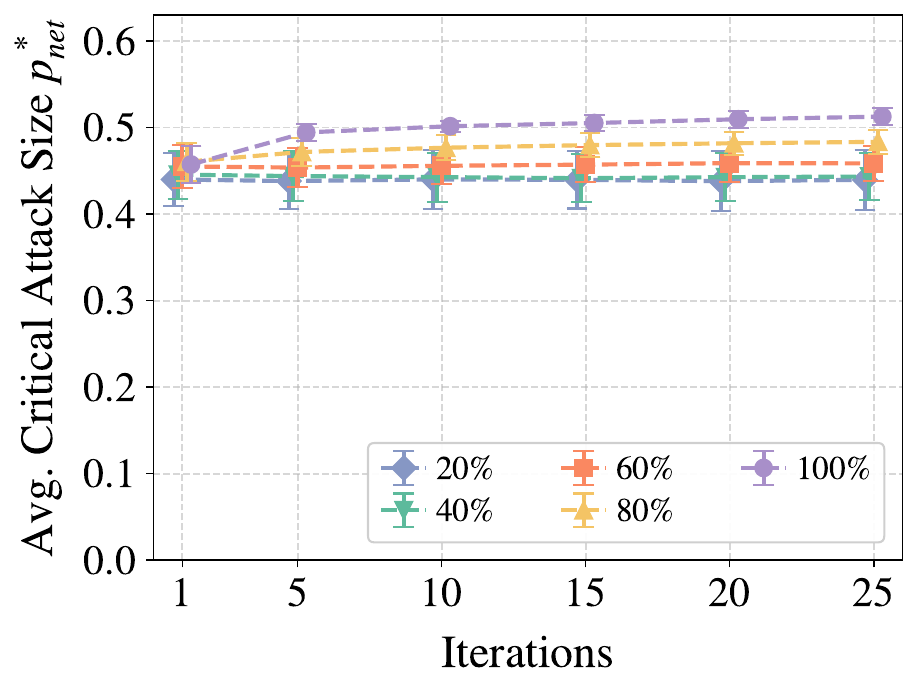}
            \caption{Network: betweenness attack}
        \end{subfigure}

        \vspace{0.25cm}

        \begin{subfigure}[b]{0.48\linewidth}
            \centering
            \includegraphics[width=\linewidth]{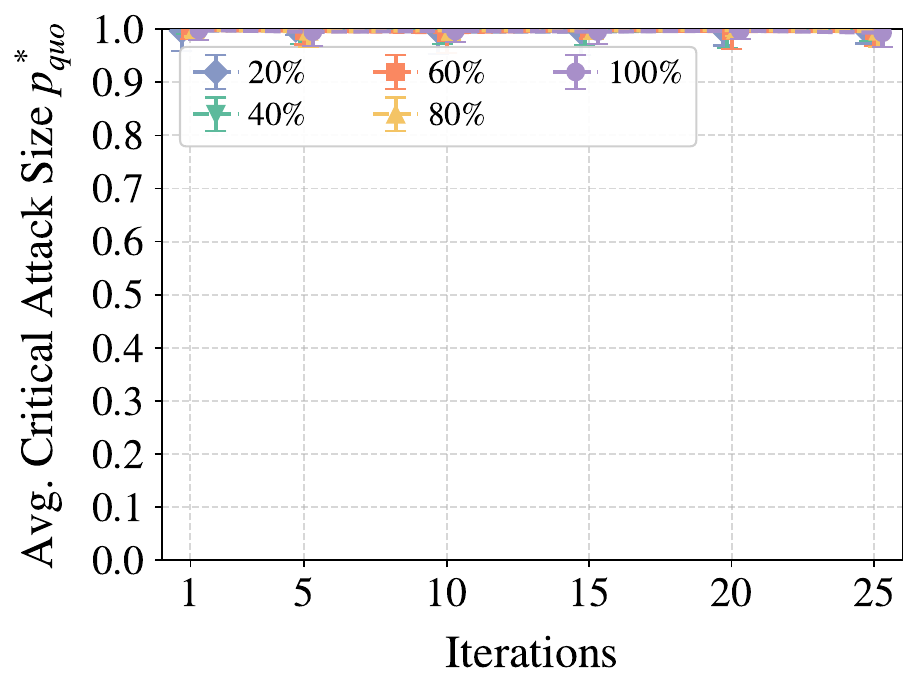}
            \caption{Quorum: degree attack}
        \end{subfigure}
        \hfill
        \begin{subfigure}[b]{0.48\linewidth}
            \centering
            \includegraphics[width=\linewidth]{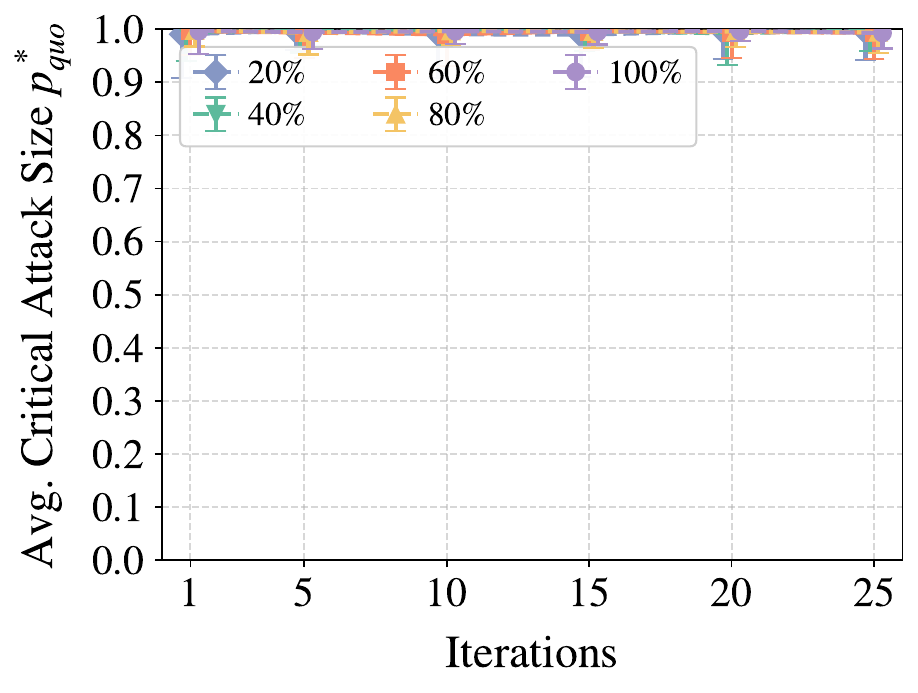}
            \caption{Quorum: betweenness attack}
        \end{subfigure}
    \end{minipage}
    \caption{Comparison of augmentation and rewiring under \textbf{degree-proportional validator selection}. Because high-degree nodes are more likely to be protected as validators, the baseline network is already more robust, leaving less room for additional gains from either mitigation strategy.}
    \label{fig:degree_prop_comparison}
\end{figure*}

\begin{figure*}
    \centering
    \begin{minipage}[t]{0.49\textwidth}
        \centering
        \textbf{Augmentation Strategy}\\[0.3em]
        \begin{subfigure}[b]{0.48\linewidth}
            \centering
            \includegraphics[width=\linewidth]{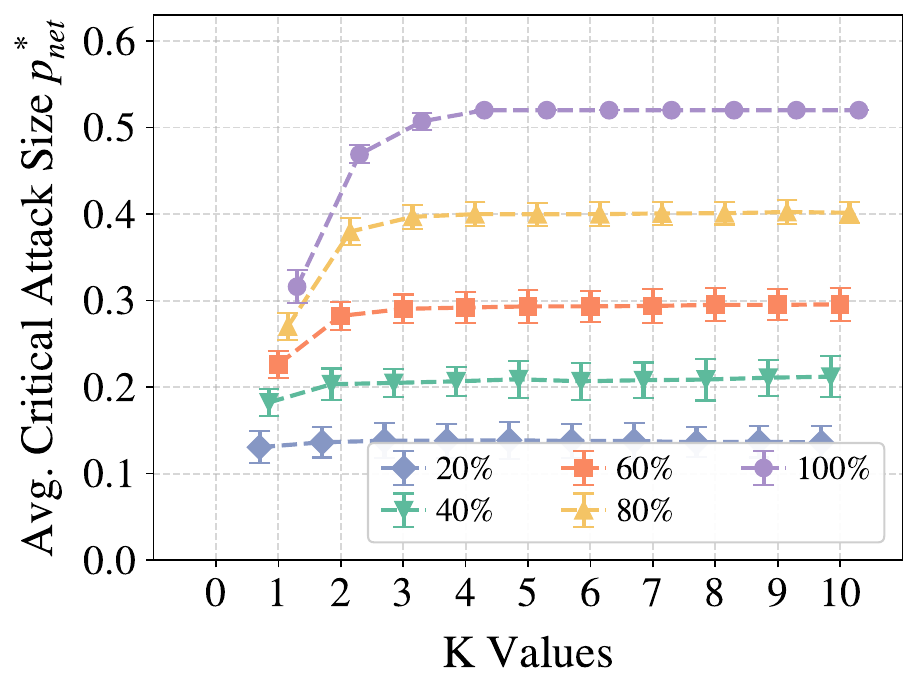}
            \caption{Network: degree attack}
        \end{subfigure}
        \hfill
        \begin{subfigure}[b]{0.48\linewidth}
            \centering
            \includegraphics[width=\linewidth]{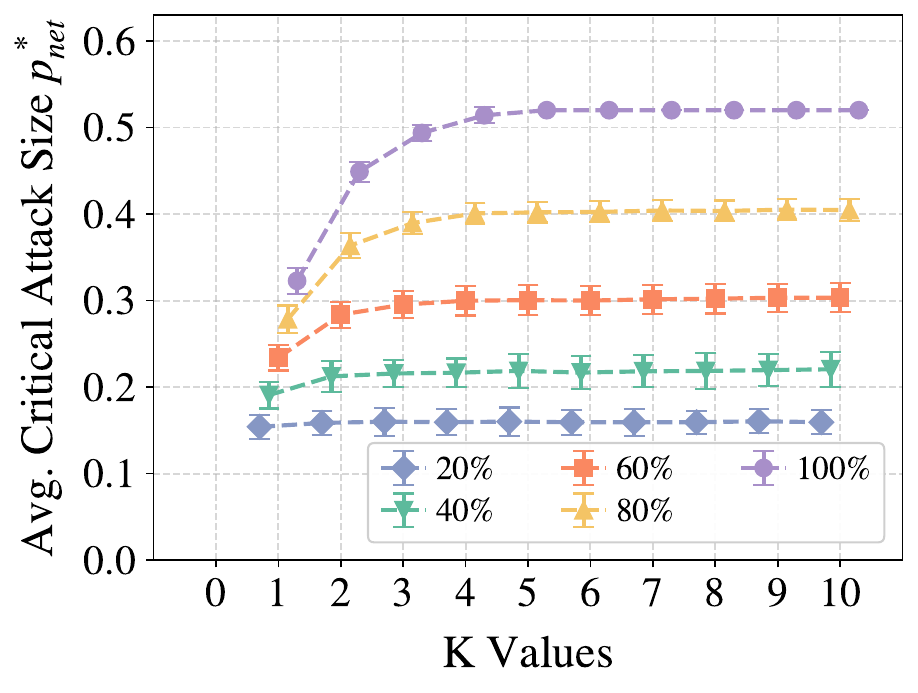}
            \caption{Network: betweenness attack}
        \end{subfigure}

        \vspace{0.25cm}

        \begin{subfigure}[b]{0.48\linewidth}
            \centering
            \includegraphics[width=\linewidth]{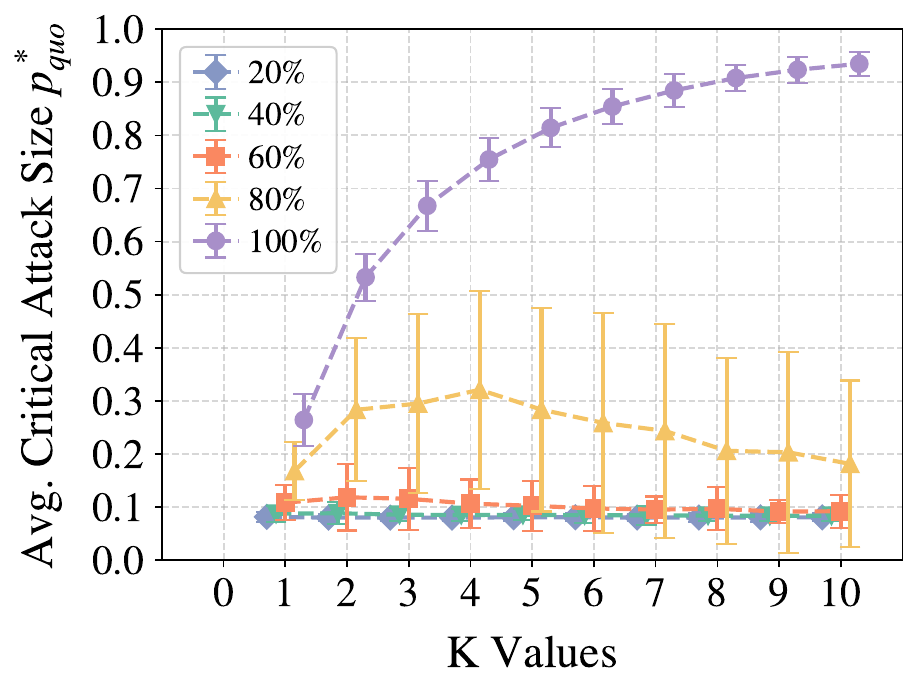}
            \caption{Quorum: degree attack}
        \end{subfigure}
        \hfill
        \begin{subfigure}[b]{0.48\linewidth}
            \centering
            \includegraphics[width=\linewidth]{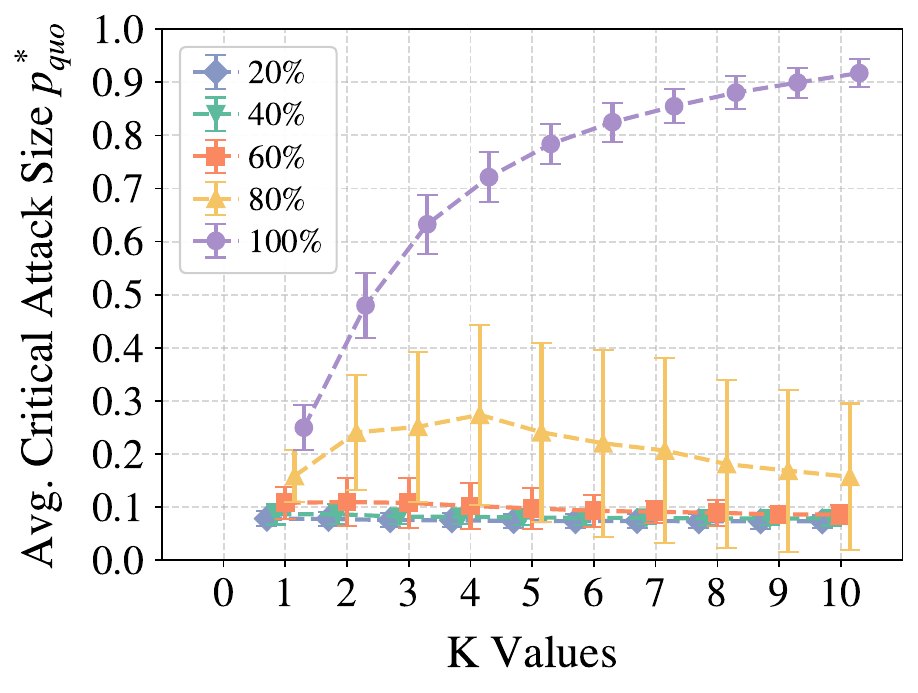}
            \caption{Quorum: betweenness attack}
        \end{subfigure}
    \end{minipage}
    \hfill
    \begin{minipage}[t]{0.49\textwidth}
        \centering
        \textbf{Rewiring Strategy}\\[0.3em]
        \begin{subfigure}[b]{0.48\linewidth}
            \centering
            \includegraphics[width=\linewidth]{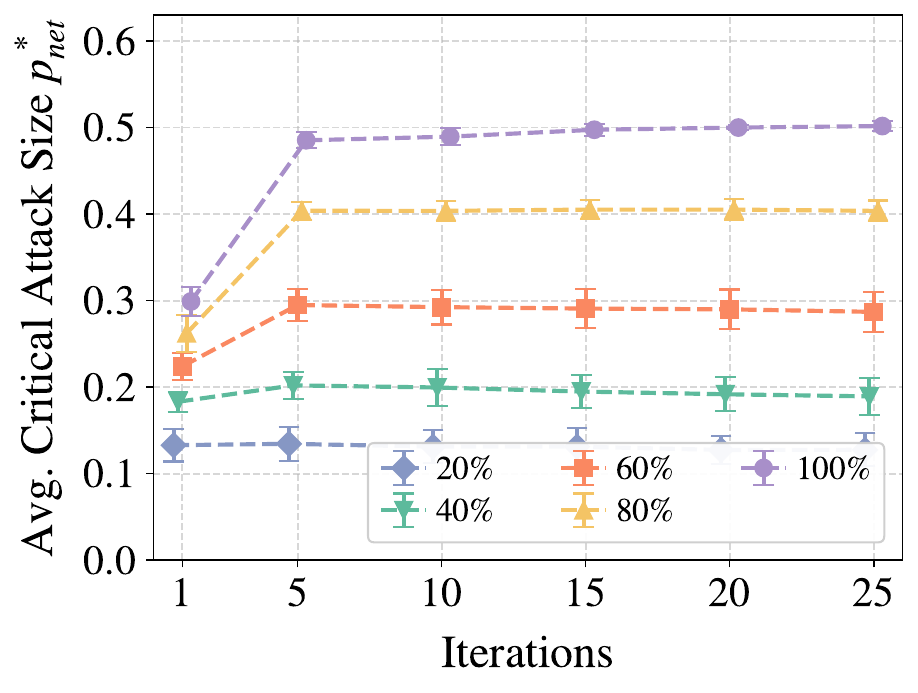}
            \caption{Network: degree attack}
        \end{subfigure}
        \hfill
        \begin{subfigure}[b]{0.48\linewidth}
            \centering
            \includegraphics[width=\linewidth]{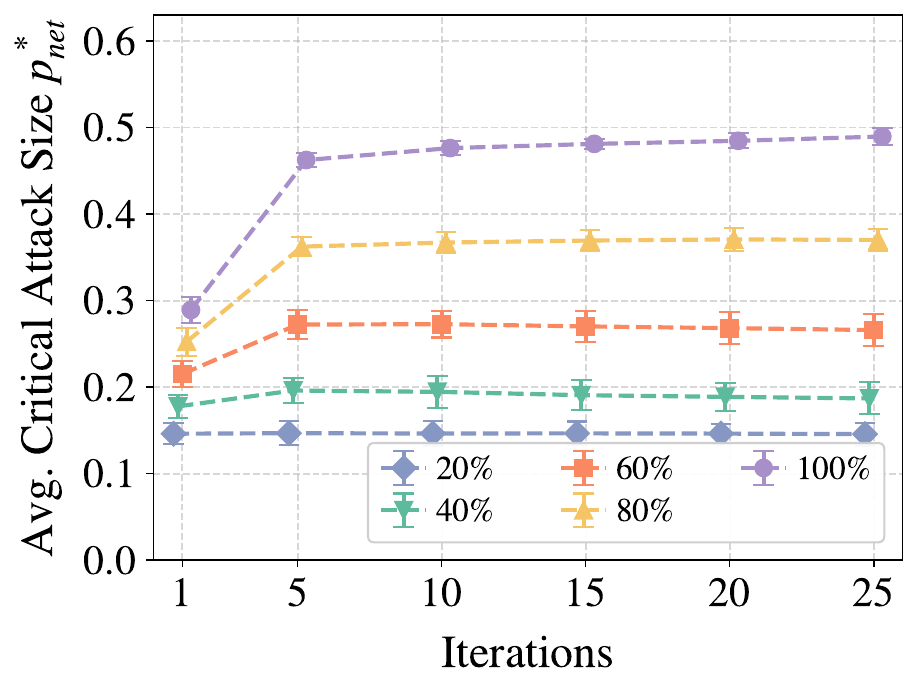}
            \caption{Network: betweenness attack}
        \end{subfigure}

        \vspace{0.25cm}

        \begin{subfigure}[b]{0.48\linewidth}
            \centering
            \includegraphics[width=\linewidth]{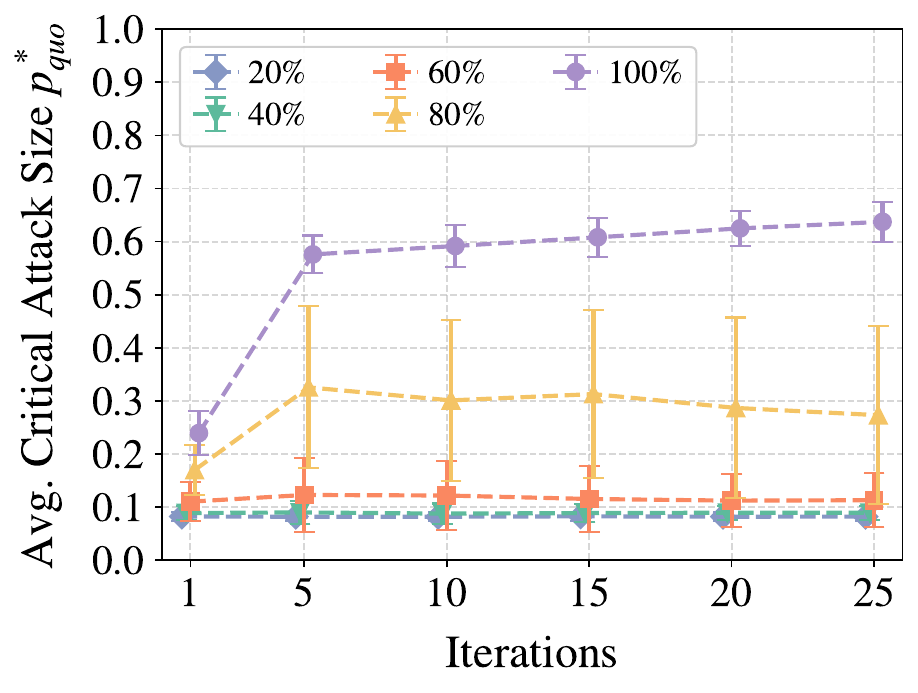}
            \caption{Quorum: degree attack}
        \end{subfigure}
        \hfill
        \begin{subfigure}[b]{0.48\linewidth}
            \centering
            \includegraphics[width=\linewidth]{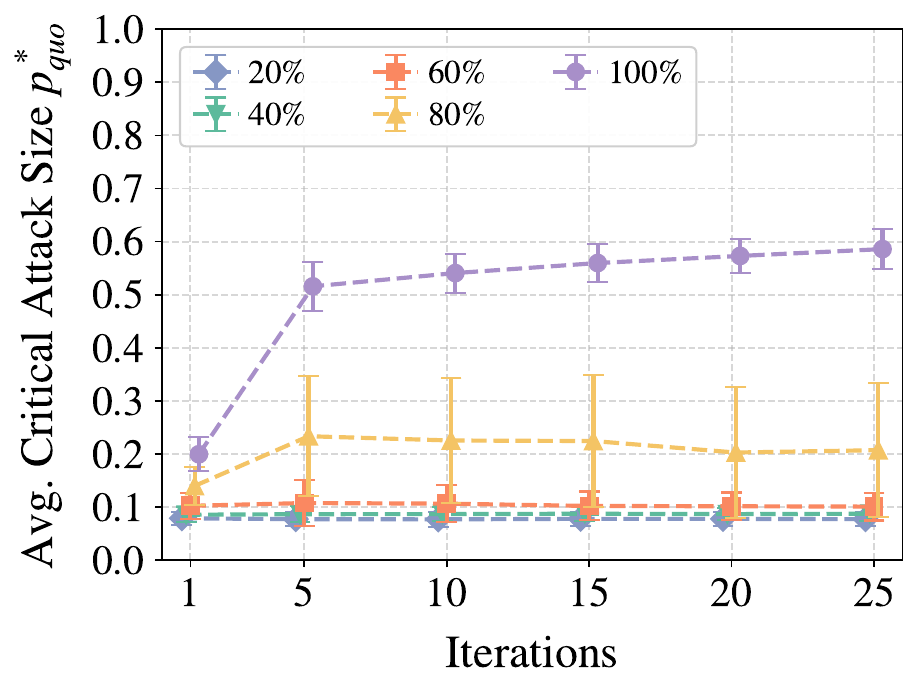}
            \caption{Quorum: betweenness attack}
        \end{subfigure}
    \end{minipage}
    \caption{Comparison of augmentation and rewiring under \textbf{inverse-degree proportional validator selection}. The qualitative pattern remains consistent with the main text: K-out augmentation matches rewiring at relatively small values of $K$ and can continue improving robustness as $K$ increases.}
    \label{fig:inverse_degree_comparison}
\end{figure*}

\section{Sensitivity to Validator-Selection Policy}
\label{appendix_1}
Because the dataset does not identify which nodes serve as validators, the validator set must be assigned synthetically in our simulations. In the main body, we examined the scenario where validators are selected uniformly at random from all nodes. Here, we present additional experiments under alternative validator selection rules to assess the sensitivity of the main conclusions. As in the main experiments, we assume that validators cannot be directly targeted.

In addition to the uniform random assignment used in the main text, we consider two alternative policies: degree-proportional selection (Figure~\ref{fig:degree_prop_comparison}), which favors high-degree nodes, and inverse-degree selection (Figure~\ref{fig:inverse_degree_comparison}), which favors low-degree nodes. 
The results are consistent with the main conclusions of the paper. 
When validators are selected proportionally to degree, the baseline network is already more robust, since many central hubs are protected as validators; consequently, both augmentation and rewiring yield smaller incremental gains. 
When validators are selected inversely proportional to degree, the baseline network is more vulnerable, yet the overall pattern remains the same: random $K$-out augmentation reaches the performance of rewiring at relatively small values of $K$, and continues to improve as $K$ increases. 
Overall, these experiments confirm that the main advantage of $K$-out augmentation, achieving the same level of robustness while preserving the original network topology, holds across different validator selection assumptions.

\end{document}